\documentclass[fleqn,12pt,twoside]{article}
\usepackage{espcrc1}

\usepackage{graphicx}
\usepackage{epsf,epsfig,psfig,float,amssymb,stmaryrd,latexsym}
\usepackage{graphics,psfrag}
\usepackage[figuresright]{rotating}

\def\3{{\ss}}

\def\vek #1 {\overrightarrow {#1}}
\newcommand{\fet}[1]{\mbox{\boldmath $#1$}}

\newcommand{\beq}{\begin{equation}}
\newcommand{\eeq}{\end{equation}}
\newcommand{\beqn}{\begin{displaymath}}
\newcommand{\eeqn}{\end{displaymath}}
\newcommand{\beqa}{\begin{eqnarray}}
\newcommand{\eeqa}{\end{eqnarray}}
\newcommand{\beqan}{\begin{eqnarray*}}
\newcommand{\eeqan}{\end{eqnarray*}}

\newcommand{\bma}{\begin{array}{cc}}
\newcommand{\ema}{\end{array}}

\newcommand{\AmS}{{\protect\the\textfont2
  A\kern-.1667em\lower.5ex\hbox{M}\kern-.125emS}}
\newcommand{\ffat}[1]{\mbox {\boldmath $#1$}}

\title{Modern theory of nuclear forces}

\author{Ulf-G. Mei{\ss}ner\address{Universit\"at Bonn, Helmholtz-Institut f{\"u}r
  Strahlen- und Kernphysik (Th) \\ D-53115 Bonn, Germany}
  \address{Forschungszentrum J\"ulich, Institut f\"ur Kernphysik (Th)\\
  D-52425 J\"ulich, Germany }}

\begin{document}

\maketitle

\begin{abstract}
Nuclear forces can be systematically derived using  effective chiral Lagrangians
consistent with the symmetries of QCD. I review the status of the calculations for
two- and three-nucleon forces and their applications in few-nucleon systems.
I also address issues like the quark mass dependence of the nuclear forces and
resonance saturation for four-nucleon operators.
\end{abstract}

\section{INTRODUCTION}
The forces between two (a few) nucleons is one of the most studied problems
in theoretical physics. Over many decades, a standard picture had evolved,
in which these forces are described in terms of meson exchanges (which is
an extension of the range expansion suggested in the 1950ties by Taketani
and collaborators \cite{range}). Nuclei, which comprise most of the baryonic 
matter in the universe, are made of slowly moving nucleons, so that given a potential,
one simply has to solve the Schr\"odinger equation for the $A$-nucleon system,
\begin{equation}
H \psi_A = E \psi_A
\end{equation}
to obtain the properties of nuclei. More precisely, the nuclear Hamiltonian
can be written as
\begin{eqnarray}
H = T + V~, \quad V = \sum_{i\neq j} V_{ij} +  \sum_{i\neq j\neq k} V_{ijk}+\ldots~,
\end{eqnarray}
with $T$ the kinetic energy operator and the 
potential $V$ is a string of terms comprising two- and three--nucleon
forces. Given the two-nucleon potential $V_{ij}$ obtained e.g. from 
the analysis of the many  $NN$ scattering
data and adjusting a few parameters in the three-nucleon potential $V_{ijk}$,
the spectra for nuclei up to $A \simeq 12$ (and other properties) 
can be calculated with high
precision using Monte-Carlo methods leading to an astonishing agreement with
data (see e.g. \cite{UAI,CSrev}). Other precision methods exist for few--nucleon systems,
like  e.g. exact solutions of the Faddeev-Yakubovsky equations, the stochastic
variational method or the use of hyperspherical harmonics.
However, in such a framework based on meson exchange or semi-phenomenological
potentials, one can not explain
why the $2N$ forces are so much stronger than the $3N$ ones and also, when
including external sources, gauge invariance is not easy to obey (but there
are prescriptions to do so). Furthermore, the connection to the fundamental theory of
the strong interactions, QCD, is loose and it is difficult to estimate the
theoretical errors. On the other hand, over the last few
decades Chiral Perturbation Theory (CHPT) has become a 
standard tool for analyzing the 
properties of hadronic systems at low energy where the perturbative 
expansion of QCD in powers of the coupling 
constant cannot be applied.  It is based on the approximate and 
spontaneously broken chiral symmetry of QCD. Starting from the most 
general effective Lagrangian for Goldstone bosons (pions in the two--flavor
case of $u$ and $d$ quarks) and matter fields (nucleons,  $\ldots$)
consistent with the symmetries of QCD,
the hadronic S--matrix elements are obtained via a simultaneous expansion
in small external momenta and quark masses (where small refers to the scale of
symmetry breaking, $\Lambda_\chi \simeq1\,$GeV). Goldstone boson loops are 
incorporated to obey perturbative unitarity and all corresponding ultraviolet 
divergences can be absorbed at a given order in the chiral expansion by the 
counterterms of the effective Lagrangian.
This perturbative scheme works well in the pion and the pion--nucleon  sectors, 
where the interaction vanishes when the external momenta  go to  zero 
(in the chiral limit). In the case of a few interacting nucleons, the
situation is very different in that one 
has to deal with  a non--perturbative system. Indeed, perturbation theory is 
expected to fail already at low energy due to the presence of the shallow 
few--nucleon bound states. To make this more precise, observe that the $np$ 
scattering length in the $^1S_0$ partial wave is much bigger than the largest
natural scale set by the pion Compton wavelength, $|a(^1S_0)| \simeq 24\,{\rm
  fm} \gg 1/M_\pi\simeq 1.4\,$fm  or consider the binding momentum in the deuteron,
which is much smaller than the pion mass, $\gamma = \sqrt{B_d m_d/2} \simeq 45\,
{\rm MeV} \ll M_\pi \simeq 140\,$MeV. A suitable non--perturbative approach has been 
suggested by Weinberg \cite{Weinb}, who showed that the strong 
enhancement of the few--nucleon  scattering amplitude arises from purely 
nucleonic intermediate states.  Weinberg suggested to apply power counting 
to the kernel 
of the corresponding scattering   
equation, which can be viewed as an effective nuclear potential. 
This idea has been explored in the last decade by many authors.
In the following I will show how far this program has matured in the 
description of few--nucleon systems. Space does not allow for giving an 
accurate representation of the historical developments and also I will
not discuss the pionless effective field theory (EFT) here (for 
comprehensive reviews see \cite{border,paulobira}).

\section{EFFECTIVE CHIRAL LAGRANGIAN}

Here I briefly discuss the effective chiral Lagrangian ${\cal L}_{\rm eff}$ 
underlying the EFT calculation of the nuclear forces. The QCD Lagrangian for 
massless up and
down quarks is invariant under global flavor $SU(2)_L \times SU(2)_R$ 
transformations or,
equivalently, under vector and axial-vector transformations. This is
called chiral symmetry. Among other facts, the absence of parity
doublets of low mass hadrons suggests that the axial symmetry is
spontaneously broken. The pions are the natural candidates for the required
Goldstone bosons. They acquire a non-vanishing mass due to the explicit
symmetry breaking caused by the small up and down quark masses. We are
interested in low-energy nuclear physics here, where the degrees of freedom are
the composite hadrons. Their interaction can described by an
effective Lagrangian, which has to be constrained by chiral symmetry and
should include explicitly symmetry breaking parts proportional to powers of
the quark masses (for the relation between the effective chiral Lagrangian
and QCD, see \cite{Heiri,SWMIT}). For the applications to be considered,
nucleon momenta are comparable to the pion mass and somewhat higher, but
still smaller than the $\rho$--mass. In that case a standard one--boson exchange
picture turns into NN contact forces for the heavy meson exchanges
and only the few-pion exchanges are kept explicitly. The
construction of the most general effective Lagrangian out of pion and
nucleon fields constrained by chiral symmetry is by now standard, based on
the non-linear realization of chiral symmetry. There is an infinite number of
possible terms, which can be ordered according to the parameter
\begin{equation}
\Delta = d + \frac{1}{2}n -2
\end{equation}
characterizing the vertices (note this differs from the standard counting done
in the pion and pion-nucleon sectors). Here $d$ is the number of derivatives and $n$
the number of nucleon field operators. Spontaneously broken chiral symmetry 
enforces $\Delta \ge 0$.
The first few terms for the interacting effective Lagrangian after a $p/m$
expansion take the form (for an early detailed review on the construction
principles and applications, see \cite{BKMrev})
\begin{eqnarray}
\mathcal{L}_{\rm eff}^{(0)} &=&  
-N^{\dagger} \left[ \frac{g_A}{2F_\pi}{\ffat \tau} {\ffat \sigma} \cdot 
\cdot {\ffat \nabla} {\ffat \pi} + \frac{1}{4F_\pi^2}{\ffat \tau} 
\cdot ({\ffat \pi} \times \dot{{\ffat \pi}}) + \cdots \right] N\nonumber\\
&&-\frac{1}{2}C_S\left(N^{\dagger}N\right)\left(N^{\dagger}N\right) 
- \frac{1}{2}C_T\left(N^{\dagger}{\ffat \sigma}N\right)
\left(N^{\dagger}{\ffat \sigma}N\right)~,\nonumber \\
\mathcal{L}_{\rm eff}^{(1)} &=& 
+\frac{1}{F_\pi^2}N^{\dagger}\left[-2c_1M_{\pi}^2{\fet \pi}^2
+c_3\partial_{\mu}{\ffat \pi}\partial^{\mu}{\ffat \pi} \right.
-\left.\frac{1}{2}c_4\varepsilon_{ijk}\varepsilon_{abc}\sigma_i\tau_a(\nabla_j\pi_b)
(\nabla_k\pi_c) + \cdots \right]N\\
&&-\frac{D}{4F_\pi}(N^{\dagger}N)(N^{\dagger}{\ffat \sigma}\cdot{\ffat \tau}N) 
\cdot \cdot {\ffat \nabla}{\ffat \pi}
-\frac{1}{2} E \,(N^{\dagger}N)(N^{\dagger}{\ffat \tau}N)^2 ~, \\
\mathcal{L}_{\rm eff}^{(2)} &=&  
-\frac{1}{2}C_1\left[(N^{\dagger}{\ffat \nabla}N)^2 
+ ({\ffat \nabla}N^{\dagger}N)^2\right] + \cdots
+ C_7(\partial_iN^{\dagger}\sigma_l\partial_iN)(N^{\dagger}\sigma_lN)
+ \cdots  ~,
\end{eqnarray}
where the upper index refers to $\Delta =0,1$ and 2. Here, $N$ and $\fet \pi$
denote the isodoublet nucleon and the isovector pion fields, in order. The parameters
of $\mathcal{L}_{\rm eff}$, 
the so called low--energy constants (LECs), can be partitioned in several
groups: some can be determined
in the $\pi /\pi$-N system ($g_A, F_\pi,c_1,c_3,c_4$) and others from nucleonic 
systems only ($C_S,C_T,E,C_1,... C_7$). The constant $D$ also affects the
NN$\pi$ system, see \cite{HMvK}. 
All these constants are of course not determined  by
chiral symmetry, but have to be adjusted to experimental data.

\section{THE HIERARCHY OF NUCLEAR FORCES}

The strength of the EFT approach is based on its underlying power counting, which
allows to organize all possible contributions to the $N$-nucleon potential
in a systematic way. As a consequence of that, one naturally obtains a
hierarchy of nuclear forces, as was shown already by Weinberg \cite{Weinb} and
van Kolck \cite{ubi1}. Here, I will discuss how this comes about including 
corrections at next-to-next-to-next-to-leading order (N$^3$LO). 
According to chiral power counting the dominant contributions
to the effective Hamiltonian for few nucleons are of the order $(Q/\Lambda)^0$, 
where $Q\sim M_\pi$ refers to the soft scale (typical momenta involved in the 
process) and $\Lambda$ to the hard scale (the chiral symmetry breaking
scale, or the  ultraviolet cut--off to render the scattering equation
finite). These leading--order (LO) contributions turn out to be 
of  the $2N$ type, given by one--pion exchange (OPE) and two 
four--nucleon contact interactions   
without derivatives, cf. Fig.~1. The first corrections at next--to--leading order (NLO) 
are of the order $(Q/\Lambda)^2$ and still of the $2N$ type. 
They result from two--pion exchange (TPE) with the leading  (i.e. with one derivative) 
$\pi NN$ and $\pi \pi NN$ vertices and seven independent
NN contact interactions with two derivatives. The 
LECs entering the expressions for TPE at NLO are the nucleon axial--vector  
coupling $g_A$ and the pion decay constant $F_\pi$. Both LECs are measured rather 
accurately, so that the leading TPE contribution is parameter--free. On the contrary,
the LECs accompanying the four-nucleon contact operators are unknown and have
to be fixed
from a fit to low angular momentum partial waves. Thus, at LO and NLO we
have only two-nucleon forces and therefore expect more--nucleon forces to be
parametrically suppressed, in agreement with empirical information.

\begin{figure}[tb]
\begin{minipage}{6cm}
Figure 1.  The hierarchy of the nuclear forces in chiral
effective field theory. Shown are representative diagrams
(contact interactions, one-, two- and three-pion
exchanges) symbolizing the various topologies at leading,
next-to-leading, $\ldots$ order. Due to parity, $2N$
contact interactions can only appear at even orders
in $Q/\Lambda$. The two-pion exchange diagrams at 
N$^2$LO (N$^3$LO) include couplings with two (three)
derivatives (or pion mass insertions)
from the $\pi N$ effective chiral Lagrangian.
At N$^3$LO, only a few topologies are shown as indicated
by the ellipsis.
\end{minipage}
\hskip 1.7 true cm
\begin{minipage}{8cm}
\psfig{file=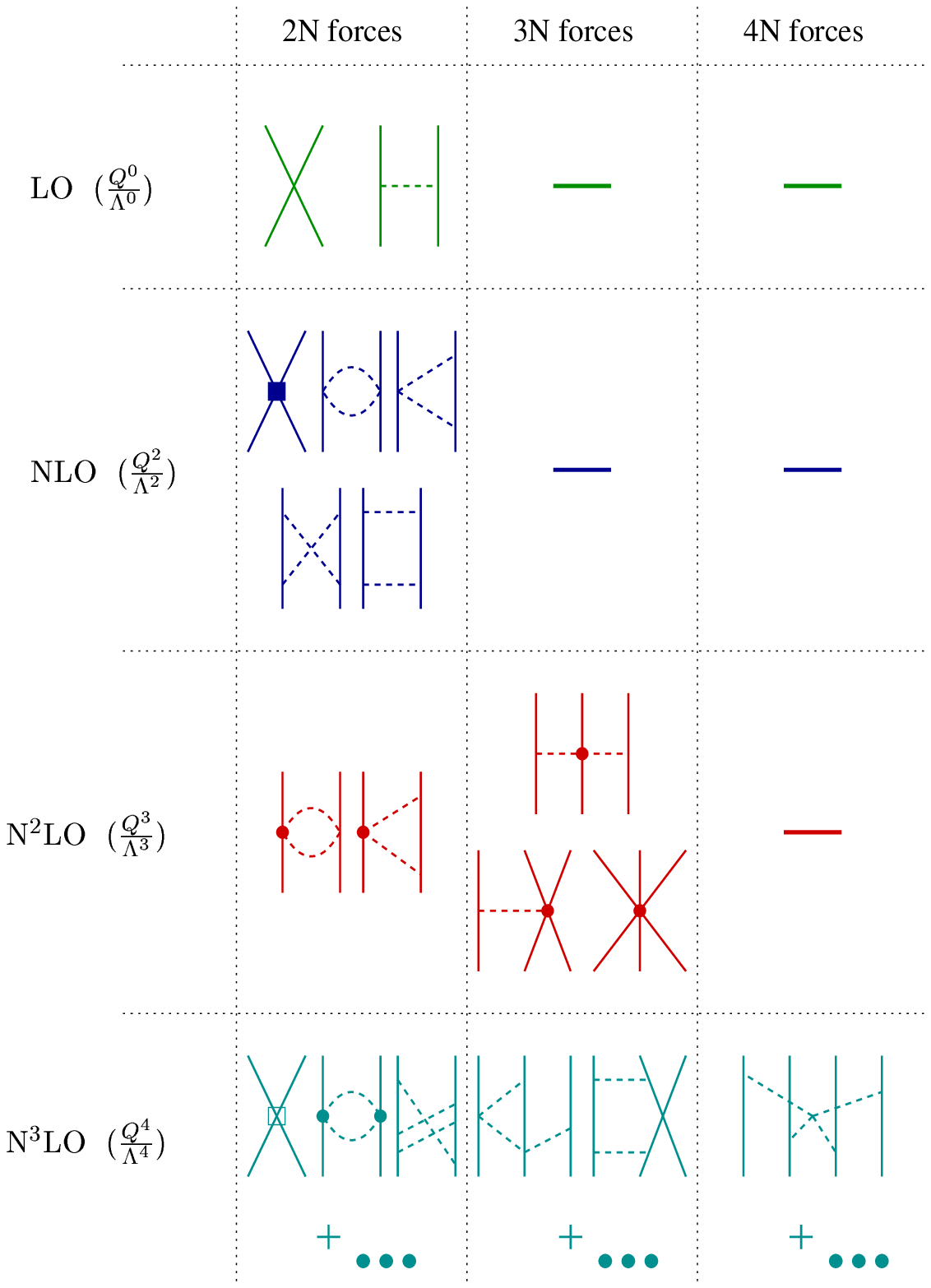,width=8cm}
\end{minipage}
\label{fig1}
\end{figure}
\setcounter{figure}{1}

\noindent
At NNLO ($\sim (Q/\Lambda)^3$) one has to take into account the subleading 
TPE contributions given by the triangle diagram with the $\pi \pi NN$ vertex 
with two derivatives or one insertion of $M_\pi^2$. The corresponding LECs
are denoted $c_{1,3,4}$ and have been fixed in the $\pi N$ system, see e.g. 
\cite{bkmlec,fms,paul00}. The numerical values of these LECs found in 
several  analyses of  $\pi N$ scattering in CHPT are rather large compared 
to what is expected on dimensional reasons. 
Similar large values of $c_{3,4}$ have also been obtained recently from 
the $np$ and $pp$ partial wave analysis carried out by the Nijmegen group 
\cite{PSA_c}. The large values of the $c_{3,4}$ can at least be partially explained 
by the fact that these LECs are saturated by the $\Delta$--excitation
\cite{bkmlec}. Other important contributions to $c_1$ and $c_4$ can be traced back
to meson--exchange in the $t$--channel. The large numerical values of the
$c_i$'s gives rise to a subleading TPE contribution to the $NN$ potential 
that shows an unphysically strong attraction already at intermediate distances 
$r \sim 1 -2$ fm when standard dimensional regularization in the pion loop 
integrals is applied. This was already pointed out in Ref.~\cite{kaiser97}. 
To circumvent this problem, spectral function regularization was proposed
in \cite{EGMspec}, as explained in the next section. It is important to note
that at this order the first non-vanishing three-nucleon forces (3NF) appear,
given by the three topologies shown in Fig.~1. I will come back to these
later. Finally, at N$^3$LO, one has to consider four--nucleon terms with
four derivatives (or pion mass insertions), further corrections to the
two--pion exchange (including now some of the dimension three couplings 
$d_i$ from ${\cal L}_{\rm eff}^{(2)}$) as well as the leading order three--pion
exchange. These TPE and 3PE corrections were worked out explicitely utilizing
DR in \cite{KaiserN3LO}. For the $NN$ problem, one has 15 independent
four--nucleon operators that feed into the S-, P- and D-waves and the
corresponding mixing angles. Further corrections to the 3NF also appear at this
order, it is however very important to stress that these are free of unknown
six-nucleon LECs. In addition, the first non-vanishing corrections to the
four--nucleon force appear at this order. Consequently, assuming that all
coefficients are of natural size, chiral symmetry applied to the few-nucleon
potential gives the hierarchy of the forces
\beq
\langle V_{2N} \rangle_A \gg \langle V_{3N} \rangle_A \gg 
\langle V_{4N} \rangle_A~, 
\eeq
which is in agreement with phenomenological determinations of these forces.

\section{SPECTRAL FUNCTION REGULARIZATION}

Before discussing the physics related to few--nucleon systems, there is one
important technical development to be discussed in a bit of detail here.
As already stated, the numerically large values of the 
LECs $c_i$ found in the $\pi N$ system lead to an unphysically strong attraction 
of subleading TPE. This is related to the use of dimensional regularization
in the pion loop integrals. This problem can be overcome employing spectral
function regularization, which has been used early in the construction of
two--pion exchange contributions to the $NN$ potential based on dispersion
theory \cite{SB}. To be specific, consider the 
isoscalar central part of the subleading TPE which results from the 
triangle diagram and is given by
\beq
\label{pot1}
V_{\rm C} (q) = \frac{3 g_A^2}{16 F_\pi^4} \int \, \frac{d^3 l}{(2 \pi)^3} 
\frac{l^2 - q^2}{((\vec q - \vec l \, )^2+ 4 M_\pi^2)((\vec q + \vec l \,)^2+ 4 M_\pi^2)} 
\left( 8 c_1 M_\pi^2 + c_3 (l^2 - q^2) \right)\,,
\eeq
where $\vec q$ is the nucleon momentum transfer and $q \equiv | \vec q \,|$, 
$l \equiv | \vec l \, |$.
The integral is cubically divergent and needs to be regularized. Applying 
dimensional regularization (DR) one finds:
\beq
\label{pot2}
V_{\rm C}  (q) = - \frac{3 g_A^2}{16 \pi F_\pi^4}
\left( 2 M_\pi^2 ( 2 c_1 - c_3 ) - c_3 q^2 \right) (2 M_\pi^2 + q^2) \frac{1}{2 q}
\arctan \frac{q}{2 M_\pi} + \ldots \;.
\eeq
The ellipses refer to polynomial  terms of the kind $\alpha + \beta
q^2$, whose explicit form  is not of relevance here.
In order to obtain the potential in coordinate space one has to make
an inverse Fourier--transform of $V_{\rm C}  (q)$ in Eq.~(\ref{pot2}). 
The ordinary  inverse Fourier--transform is obviously not possible 
due to the the fact that $V_{\rm C}  (q)$
grows with $q$. One can nevertheless obtain $V_{\rm C}  (r)$
at each $r > 0$ using the spectral function representation \cite{kaiser97}:
\beq
\label{spectrfun}
V_{\rm C}  (q) = \frac{2 q^4}{\pi} \int_{2 M_\pi}^\infty d \mu \, \frac{1}{\mu^3}
\, \frac{\rho (\mu)}{\mu^2 + q^2},
\eeq
where the spectral function $\rho (\mu )$ can be obtained from 
$V_{\rm C} (q )$ in Eq.~(\ref{pot2}) via
\beq
\label{rho}
\rho (\mu ) = \Im \left[ V_{\rm C} (0^+ - i \mu ) \right]
= - \frac{3 g_A^2}{64 F_\pi^4} \left( 2 M_\pi^2 ( 2 c_1 - c_3) + c_3 \mu^2 \right)
(2 M_\pi^2- \mu^2) \frac{1}{\mu} \theta (\mu - 2 M_\pi )\,.
\eeq
In eq.~(\ref{spectrfun}) a twice subtracted dispersion integral
is given which is needed in order to account for the large--$\mu$
behavior of $\rho (\mu)$.

\begin{figure}[tb]
\centerline{
\psfig{file=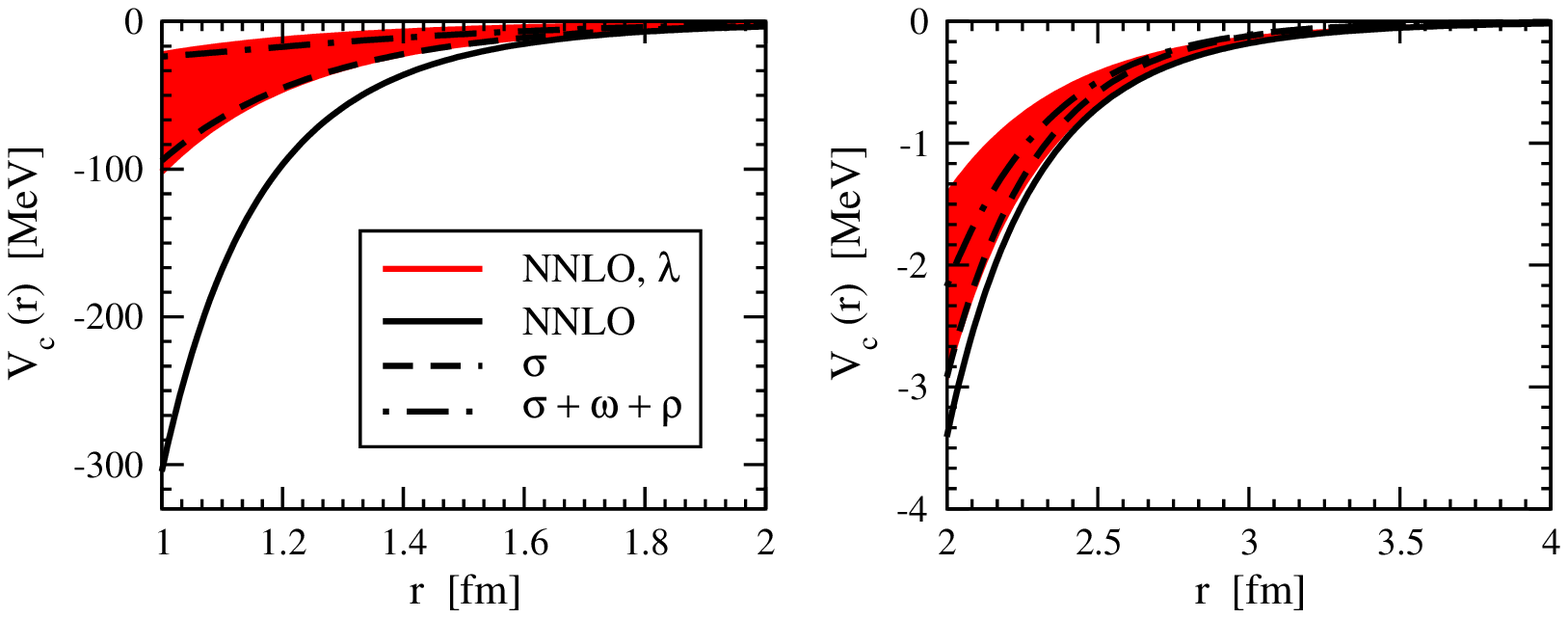,width=12.5cm}}
\vskip -0.7 true cm
\centerline{
\parbox{1.00\textwidth}{
\caption[fig2]{
\label{fig2}  The potential $V_C$  in $r$--space. 
The solid line (shaded band) shows the DR (spectral function regularized, 
$\lambda = 500 \ldots 800$ MeV) result. The dashed (dashed--dotted) line 
refers to the phenomenological $\sigma$ ($\sigma + \omega + \rho$) contributions based on the
isospin triplet configuration space version (OBEPR) of the Bonn potential.   
}}}
\end{figure}

\noindent
The inverse Fourier--transform in terms of the spectral function 
$\rho (\mu )$ can easily be evaluated via
\beq
\label{four}
V_{\rm C} (r ) = \frac{1}{2 \pi^2 r} \int_{2 M_\pi}^\infty d \mu
\, \mu \, e^{- \mu r} \rho (\mu ).
\eeq
Substituting $\rho (\mu )$ from Eq.~(\ref{rho}) into Eq.~(\ref{four})
and using for the LECs $c_{1,3}$ the central values from \cite{paul00},
$c_1 = -0.81$ GeV$^{-1}$ and $c_3 = -4.70$ GeV$^{-1}$,
one obtains the coordinate space representation of the potential
$V_{\rm C} (r )$ shown by the solid line 
in Fig.~\ref{fig2}. The central part of the NNLO TPE
potential turns out to be several times more attractive 
at intermediate distances than the 
phenomenological $\sigma$ ($\sigma + \omega + \rho$) contributions. 
This unphysical attraction shows up in the D-- and F--wave phase shifts 
as  observed in \cite{kaiser97}.
The origin of the unphysical attraction at NNLO 
can be traced back by looking at the integral in Eq.~(\ref{four}).  
While at large distances the integral is dominated by  low--$\mu$ components
(of the order $\mu \sim 350$ MeV), already at intermediate 
distances  rather high--$\mu$ terms (of the order  $\mu \sim 600$ MeV) 
in the spectral function provide a dominant contribution. 
Clearly, at shorter distances even higher--$\mu$ components become 
important. Chiral EFT can hardly provide convergent results 
for the spectral function at $\mu \sim 600$ MeV and higher. Instead 
of keeping such large--$\mu$ contributions in the regularized loop integral 
expressions it is advantageous to perform the integration in 
Eqs.~(\ref{spectrfun},\ref{four}) 
only over the low--$\mu$ region, where chiral EFT is applicable. 
This can be achieved by introducing the regularized spectral function 
\beq
\label{regspectr}
\rho (\mu ) \rightarrow \rho^\lambda (\mu ) = \rho (\mu ) \, \theta (\lambda - \mu )\,,
\eeq
with the reasonably chosen cut--off $\lambda < M_\rho$. 
Certainly, taking a too small $\lambda$ in Eq.~(\ref{regspectr}) will remove 
the truly long--distance physics while too large values for the cut--off 
may affect the convergence of the EFT expansion due to the inclusion of spurious 
short--distance physics.  In Fig.~\ref{fig2} we show $V_{\rm C} (r)$ obtained 
using the spectral function regularization Eq.~(\ref{regspectr}) with 
$\lambda = 500 \ldots 800$ MeV.  The strongest effects of the cut--off are 
observed at intermediate and short distances, where the unphysical attraction 
in dimensionally regularized TPE is greatly reduced. On the other hand, 
the asymptotic behavior of the potential at large $r$ is not affected by the
choice of regularization. The results for D-- and F--waves are greatly improved 
when the spectral function regularization is used instead of DR, see \cite{EGMspec}.
One can also show  that the spectral function 
regularization is equivalent to (finite) momentum cut--off regularization 
of pion loop integrals. 
It should be understood that this new regularization scheme does 
not introduce any model dependence in the EFT procedure as soon as 
$\lambda$ is chosen of the order of (or larger than) $M_\rho$. 
Various choices for $\lambda$ (including $\lambda = \infty$, which is equivalent
to DR) differ from each other by higher--order contact
terms and lead to exactly the same result for observables provided one 
keeps terms in all orders in the EFT expansion. Of course, this
choice of regularization in Eq.~(\ref{regspectr})
is by no means unique. Different choices lead to equivalent results for the 
potential up to higher order terms and may be used as well. The advantage 
of the form  Eq.~(\ref{regspectr}) is that it does not generate spurious 
long--range contributions which are suppressed by inverse powers of
$\lambda$.  For a more detailed discussion on this issue, I refer to \cite{EGMspec}.

\section{TWO NUCLEONS AT N$^3$LO}

\begin{figure}[htb]
\centerline{
\psfig{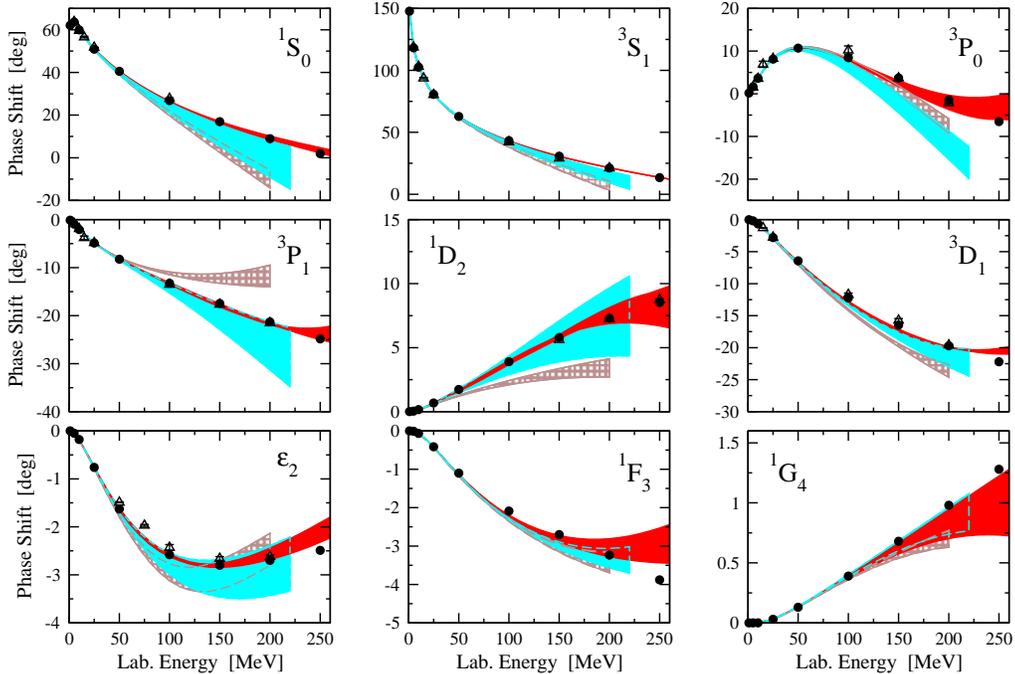}
}
\vskip -0.7 true cm
\centerline{
\parbox{1.00\textwidth}{
\caption[fig3]{
\label{fig3}
Selected  {\it np} phase shifts  and  mixing angle $\epsilon_2$
 versus the nucleon laboratory
energy. The grid, light shaded and dark shaded bands show the NLO, 
NNLO \cite{EGMspec} and N$^3$LO results, respectively.
The filled circles depict the Nijmegen PWA results \cite{nijpwa}
and the open triangles are the results from the  Virginia Tech PWA \cite{said}. 
}}}
\vspace{-0.3cm}
\end{figure}

\begin{figure}[tb]
\centerline{
\psfig{file=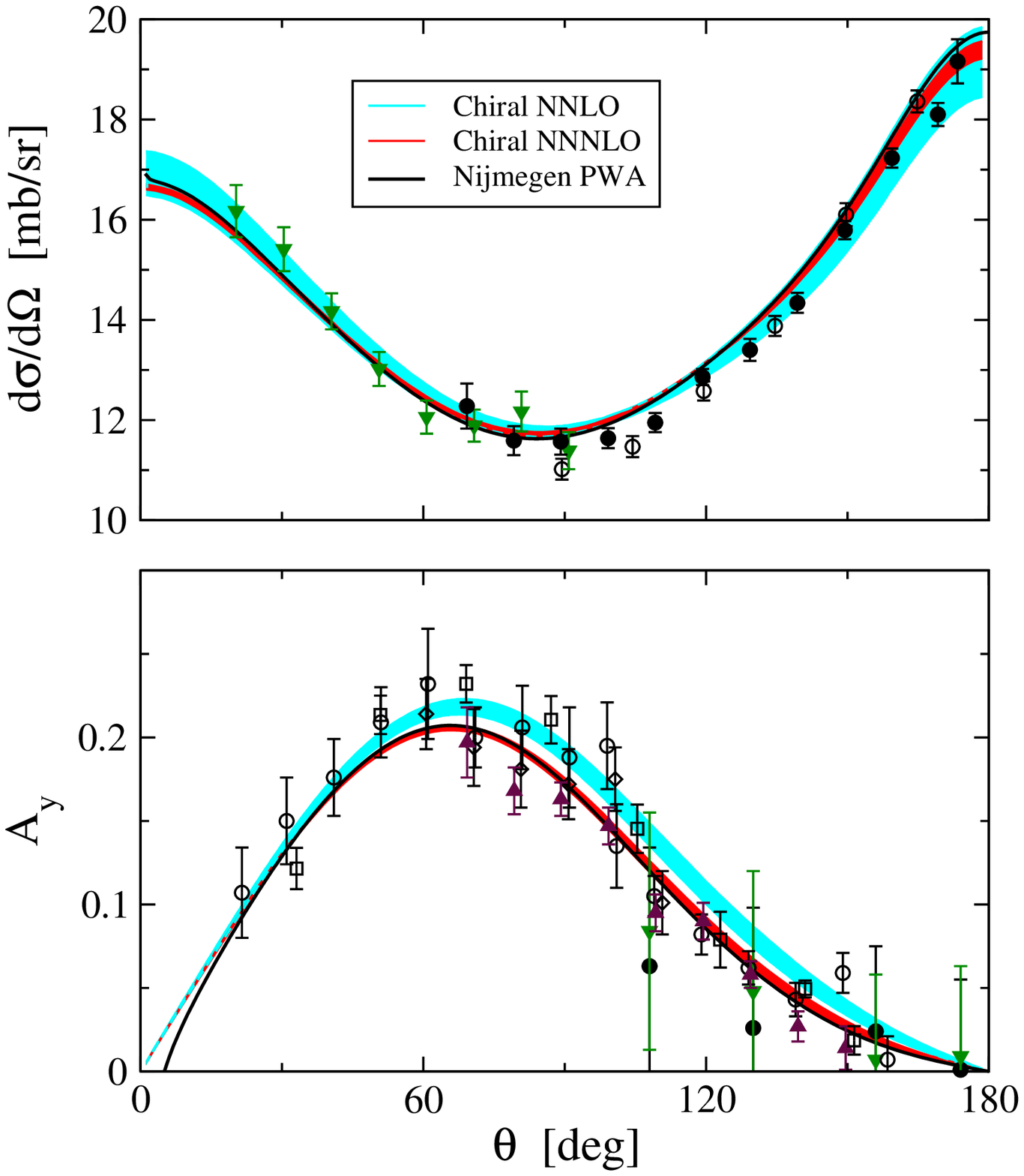,width=7.2cm}
\psfig{file=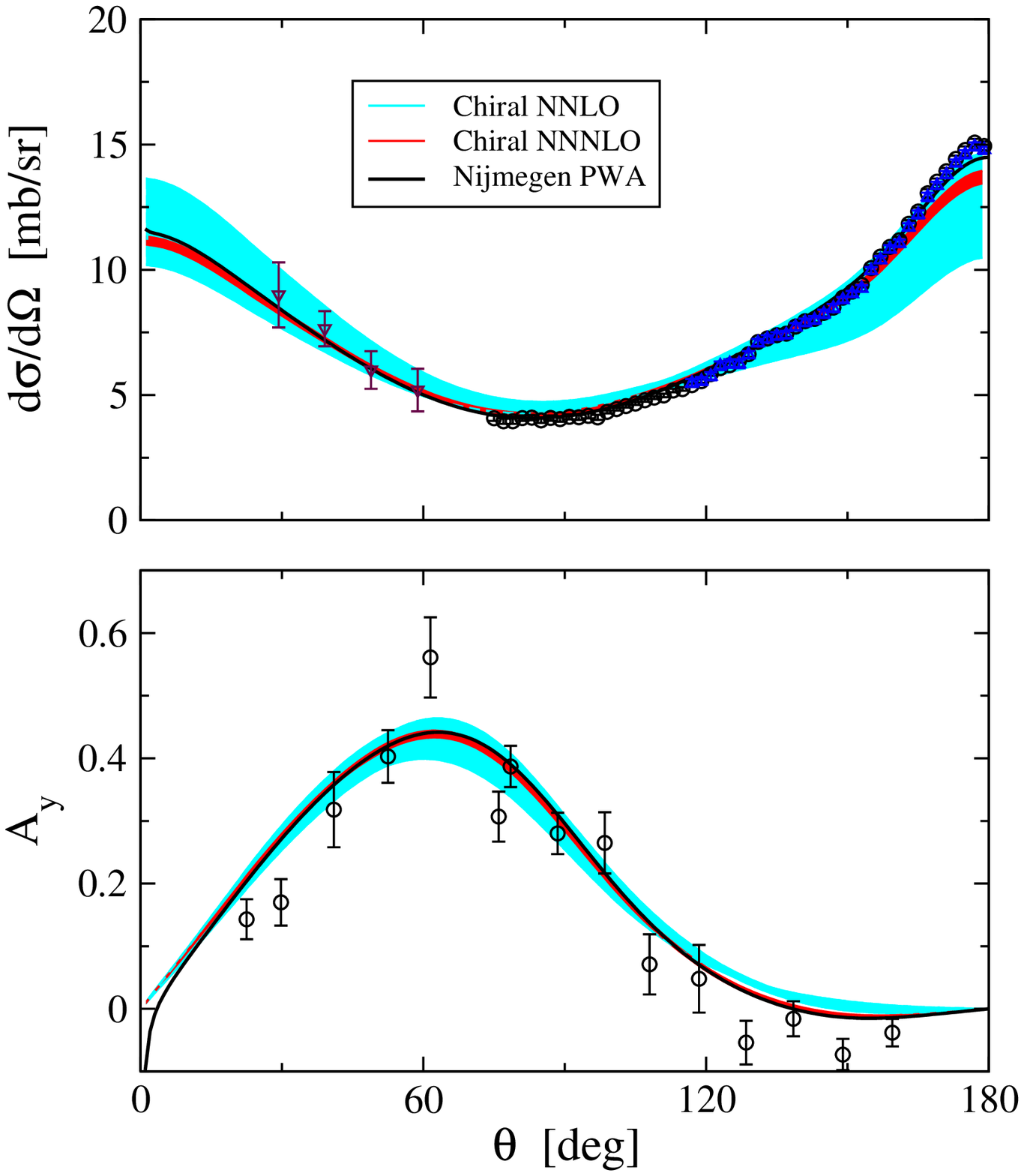,width=7.2cm}
}
\vskip -0.7 true cm
\centerline{
\parbox{1.00\textwidth}{
\caption[fig4]{
\label{fig4}  
{\it np} differential cross section (upper row) and vector analyzing power 
(lower row) at 
$E_{\rm lab}=50$ MeV (left panel) and $E_{\rm lab}=96$ MeV (right panel).
Shaded bands refer to the NNLO result, dashed lines to the Nijmegen phase shift analysis
(NPSA) \cite{nijpwa}. 
}}}
\end{figure}

The  interactions between two nucleons at N$^3$LO were studied in
\cite{EGMn3lo}. Here, I can only sketch the most salient features of that
study. As discussed before, at this order
the two--nucleon potential consists of one-, two- and three-pion
exchanges and a set of contact interactions with zero, two and four
derivatives, respectively. We have applied spectral function regularization to the
multi-pion exchange contributions.  Within this framework, 
we have shown that three-pion exchange can safely be neglected. The
corresponding cut--off is varied from 500 to 700 MeV. The LECs 
related to the dimension two and three $\bar NN\pi\pi$ 
vertices are taken consistently from studies of pion-nucleon scattering in
chiral perturbation theory \cite{fms,paul00}.
In the isospin limit, there are 24 LECs related to four--nucleon 
interactions which feed into the
S--, P-- and D--waves and various mixing parameters.
In addition,  various isospin breaking mechanisms were considered.
In the actual calculations, we have
included the leading charge-independence and charge-symmetry breaking
four--nucleon operators, the pion and nucleon mass differences in the 1PE, 
and the same  electromagnetic corrections as done by
the Nijmegen group  (the static Coulomb potential and various corrections to
it, magnetic moment interactions and vacuum polarization). This is done because
we fit to the Nijmegen partial waves. In the future, it would be important to
also include mass differences in the 2PE, $\pi\gamma$-exchange and the isospin
breaking corrections to the pion-nucleon scattering amplitude (which have been
consistently determined in \cite{fet01}). We therefore have phases for the
$pp$, $np$ and $nn$ systems.
To obtain the bound and scattering states, we use
the Lippmann-Schwinger equation with the relativistic form of the kinetic
energy. Such an approach can easily be extended to external probes or
few--nucleon systems.
The LS equation is regulated in the standard way, namely by
multiplying the potential $V (\vec p, \; \vec p \, ')$ with a 
regulator function  $f^\Lambda$,
\beq
\label{pot_reg}
V (\vec p, \; \vec p \, ') \rightarrow f^\Lambda ( p ) \, 
V (\vec p, \; \vec p \, ')\, f^\Lambda (p ' )\,.
\eeq 
We use the exponential regulator function 
$f^\Lambda (p ' ) = \exp [- p^6/\Lambda^6 ]\,$,
with the cut-off varied from 450 to 600~MeV.
The total of 26 four--nucleon LECs has been determined by
a combined fit  to some $np$ and $pp$ phase shifts from the Nijmegen analysis
together with the $nn$ scattering length value $a_{nn} = -18.9\,$fm. 
The resulting LECs are of natural
size except $D_{1S0}^1$ and $D_{3S1}^1$. In contrast to the fits at 
NLO and NNLO, we had to extend the
fit range to $E_{\rm lab} = 200\,$MeV.
The description of the low phase shifts (S, P, D) is excellent, see
Fig.~\ref{fig3} for the S- and some selected P- and D-waves. 
In all cases, the N$^3$LO result is better 
than the NNLO one with a sizeably reduced theoretical uncertainty. This 
holds in particular for the problematic $^3P_0$ wave which was not well 
reproduced at NNLO. The peripheral waves (F, G, H, $\ldots$),
that are free of parameters, are also well described with the
expected theoretical uncertainty related to the cut--off variations, 
see Fig.~\ref{fig3} for $^1F_3$ and $^1G_4$. 
We stress that the description of the
phases in general improves when going from LO to NLO to NNLO to N$^3$LO,
as it is expected in a converging EFT. 
The resulting S-wave scattering lengths and range parameters in the $np$
and $pp$ systems 
are in good agreement with the ones obtained in the Nijmegen PWA. In addition,
we can give theoretical uncertainties for all these quantities, which are
mostly in the one percent range. 
The  scattering observables (differential cross sections,
analyzing powers) for the $np$ system as shown e.g. in  
Fig.~\ref{fig4} are well described, with a small theoretical
uncertainty at the order considered here.
The deuteron properties are further predictions. In particular,
we have not included the binding energy in the fits, the deviation from the
experimental value is in the range from 0.4 to 0.07$\%$. The asymptotic $S$-wave
normalization and the asymptotic $D/S$ ratio are also well described, see Table~1.
The remaining
discrepancies in the quadrupole moment and the rms matter radius are related
to the short-ranged two-nucleon current not considered in \cite{EGMn3lo}. 
Note that the $2N$ system at this order  was also studied in \cite{EM}, utilizing 
dimensional regularization in the pion loop graphs. In
that work, however, no detailed dicsussion of the  theoretical uncertainties was given.

\begin{table}[htb]
\caption{Deuteron properties at various orders in chiral EFT in comparison
to the data.}
\begin{tabular}{|c|c|c|c|c|}
\hline
             & NLO   & N$^2$LO   & N$^3$LO   & Exp. \\ \hline
$E_d$ [MeV]  & $-$2.171 \ldots $-$2.188 & $-$2.189 \ldots $-$2.202 &
               $-$2.216 \ldots $-$2.223 & $-$2.2246 \\ \hline
$A_S$ [fm$^{1/2}$] & $0.868 \ldots 0.873$ & $0.874 \ldots 0.879$
                   & $0.882 \ldots 0.883$ & 0.8846(9)\\ \hline
$\eta$             & 0.0256 \ldots 0.0257 & 0.0255 \ldots 0.0256 
                   & 0.0254 \ldots 0.0255 & 0.0256(4)\\ \hline 
\end{tabular}
\end{table}
\noindent

\section{THREE NUCLEONS AT N$^2$LO}

In the chiral EFT framework, three-nucleon forces (3NFs) arise consistently
with the $NN$ forces from the effective Lagrangian. This is one of the major 
advantages of this approach - such a consistency has never been achieved before.
Furthermore, chiral EFT also allows to generate the most general structures
consistent with the underlying symmetries, this is simply a consequence of 
utilizing the most general effective Lagrangian consistent with these principles.
As noted earlier, the leading non-vanishing contributions to the 3NF  appear 
at NNLO, given by three different topologies (see Fig.~1): the TPE graphs,
the OPE diagram and the six-nucleon contact interactions. The TPE contribution is
given in terms of LECs $c_{1,3,4}$ from the pion-nucleon system (see \cite{FHvK,EGM3NF}) 
\beq
\label{3nftpe}
V^{\rm 3NF}_{\rm TPE}=\sum_{i \not= j \not= k} \frac{1}{2}\left(
  \frac{g_A}{2 F_\pi} \right)^2 \frac{( \fet \sigma_i \cdot \fet q_{i}
  ) 
(\fet \sigma_j \cdot \fet q_j  )}{(\fet q_i\, ^2 + M_\pi^2 ) ( \fet
q_j\, ^2 + M_\pi^2)}  F^{\alpha \beta}_{ijk} \tau_i^\alpha 
\tau_j^\beta \,,
\eeq
where  $\fet q_i \equiv \fet p_i \, ' - \fet p_i$; $\fet p_i$
($\fet p_i \, '$) are the initial (final) momenta of the nucleon $i$ and 
\begin{displaymath}
F^{\alpha \beta}_{ijk} = \delta^{\alpha \beta} \left[ - \frac{4 c_1
    M_\pi^2}{F_\pi^2}  + \frac{2 c_3}{f_\pi^2}  
\fet q_i \cdot \fet q_j \right] + \sum_{\gamma} \frac{c_4}{F_\pi^2} \epsilon^{\alpha
\beta \gamma} \tau_k^\gamma  
\fet \sigma_k \cdot [ \fet q_i \times \fet q_j  ]\,.
\end{displaymath}
The OPE and contact contributions are given in terms of the LECs $D$ and $E$,
respectively, and take the form
\beqa
\label{3nfrest}
V^{\rm 3NF}_{\rm OPE} &=& - \sum_{i \not= j \not= k} \frac{g_A}{8
  F_\pi^2} \, D \, \frac{\fet \sigma_j \cdot \fet q_j }{\fet q_j\, ^2
  + M_\pi^2}  
\, \left( \fet \tau_i \cdot \fet \tau_j \right) 
(\fet \sigma_i \cdot \fet q_j ) \,, \\
V^{\rm 3NF}_{\rm cont} &=& \frac{1}{2} \sum_{j \not= k}  E \, 
( \fet \tau_j \cdot \fet \tau_k ) \,.
\eeqa
To pin down the 3NF at this order, one  needs two observables to determine
the LECs $E$ and $D$. In \cite{EGM3NF},  these two parameters were determined 
from the $^3$H binding energy and the $^2a_{nd}$ scattering length. In that
paper, a detailed study of 3- and 4-nucleon observables including this 3NF was
performed. 
The role of the 3NF increases with increasing energy, as exemplified
in Fig.~5, where the minimum of the differential cross section for $nd$ scattering
at 65 MeV is shown. It is clearly visible that the calculation without 3NF badly
misses the data. I should stress again that the solid curve including the 3NF
is a prediction since all parameters have been fixed before.

\begin{figure}[htb]
\begin{minipage}{6cm}
Figure 5.  Differential cross section for $nd$ scattering at
65 MeV around the minimum at $\theta \simeq 130^\circ$.
Solid (dashed) line: NNLO prediction with (without) 3NF. 
The dot-dashed line
gives the prediction based on the CD-Bonn potential without 3NF.
\end{minipage}
\hskip 1.7 true cm
\begin{minipage}{8cm}
\psfig{file=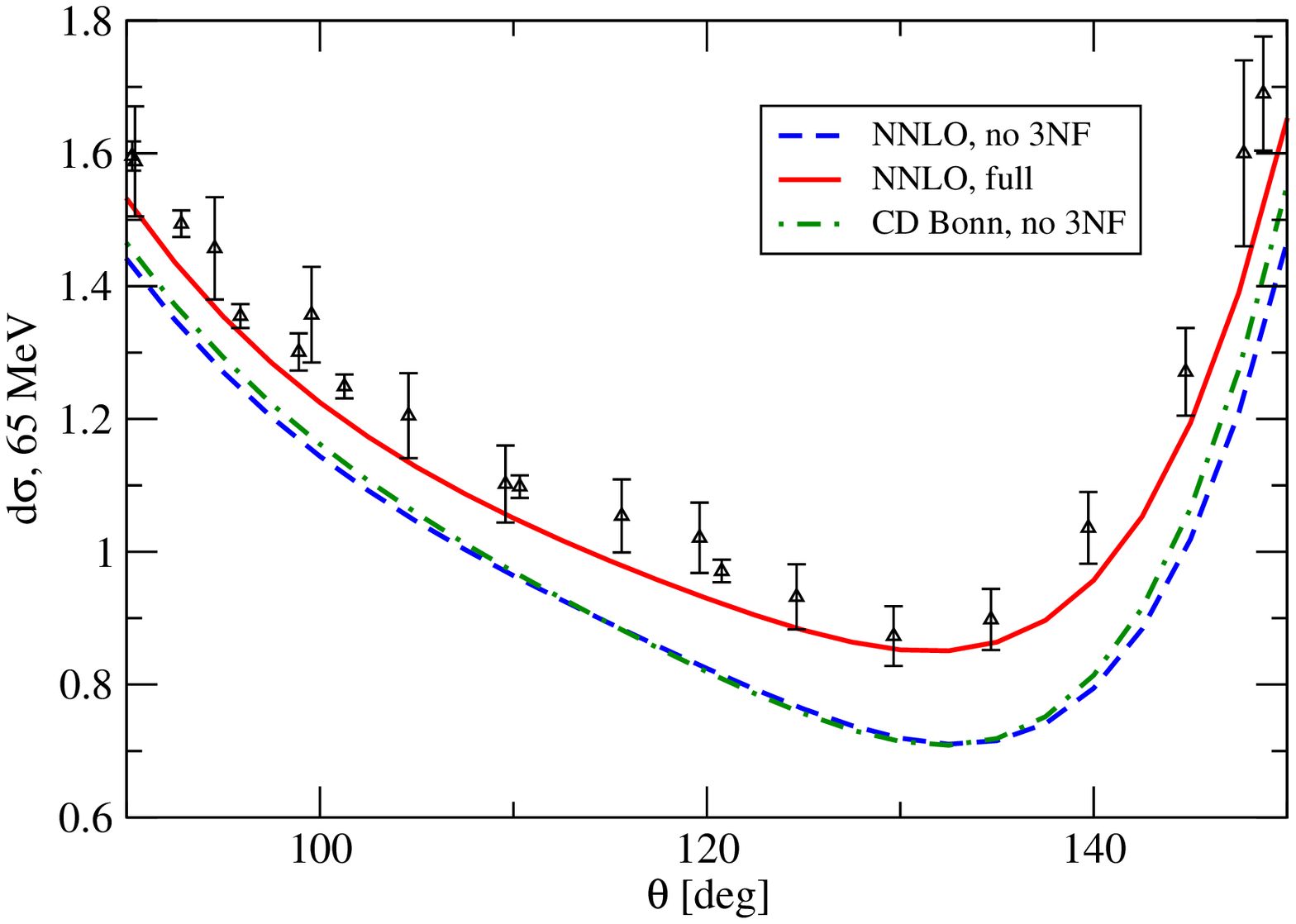,width=7cm}
\end{minipage}
\label{fig3N1}
\end{figure}
\setcounter{figure}{5}

\begin{figure}[htb]
\centerline{
\psfig{file=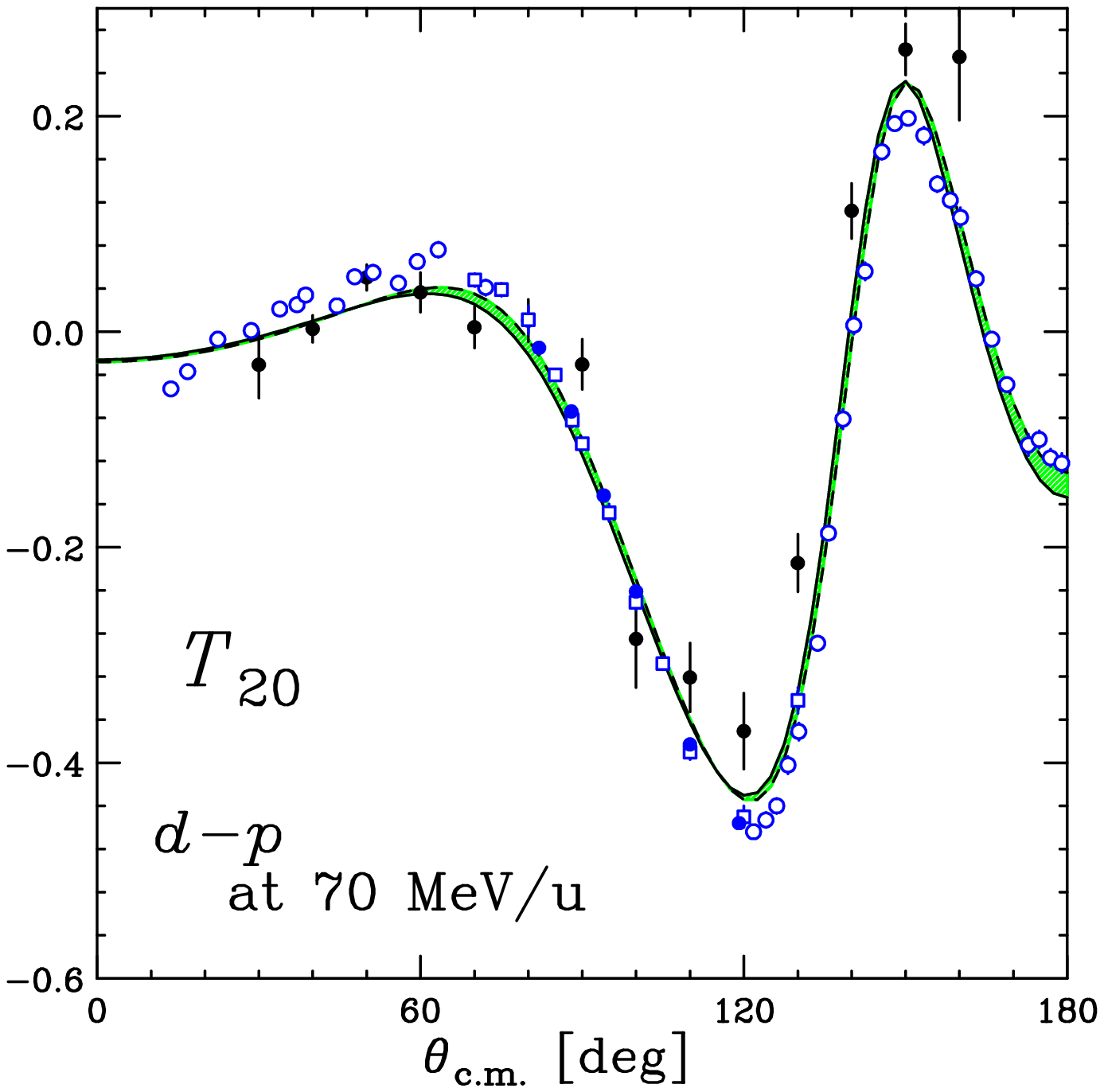,width=6.5cm}
\psfig{file=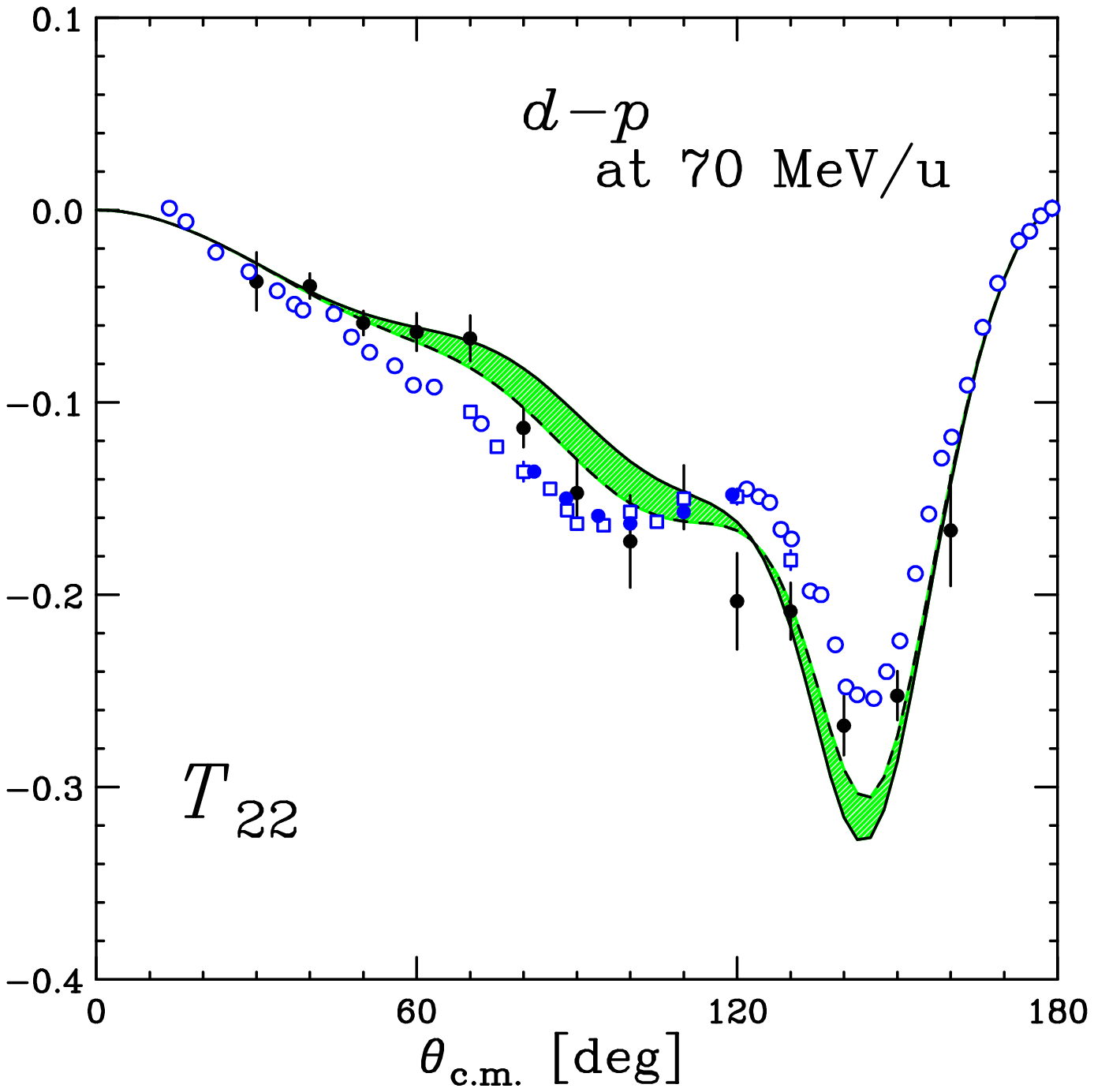,width=6.5cm}
}
\vskip -0.7 true cm
\centerline{
\parbox{1.00\textwidth}{
\caption[fig3N3]{
\label{fig3N3}  
Left: Tensor analyzing powers $T_{20}$ (left) and $T_{22}$ (right)
in  $pd$ scattering at 70~MeV recently measured at RIKEN \cite{sekipriv}.
The green band shows the NNLO calculation.
}}}
\end{figure}

\begin{figure}[htb]
\centerline{
\psfig{file=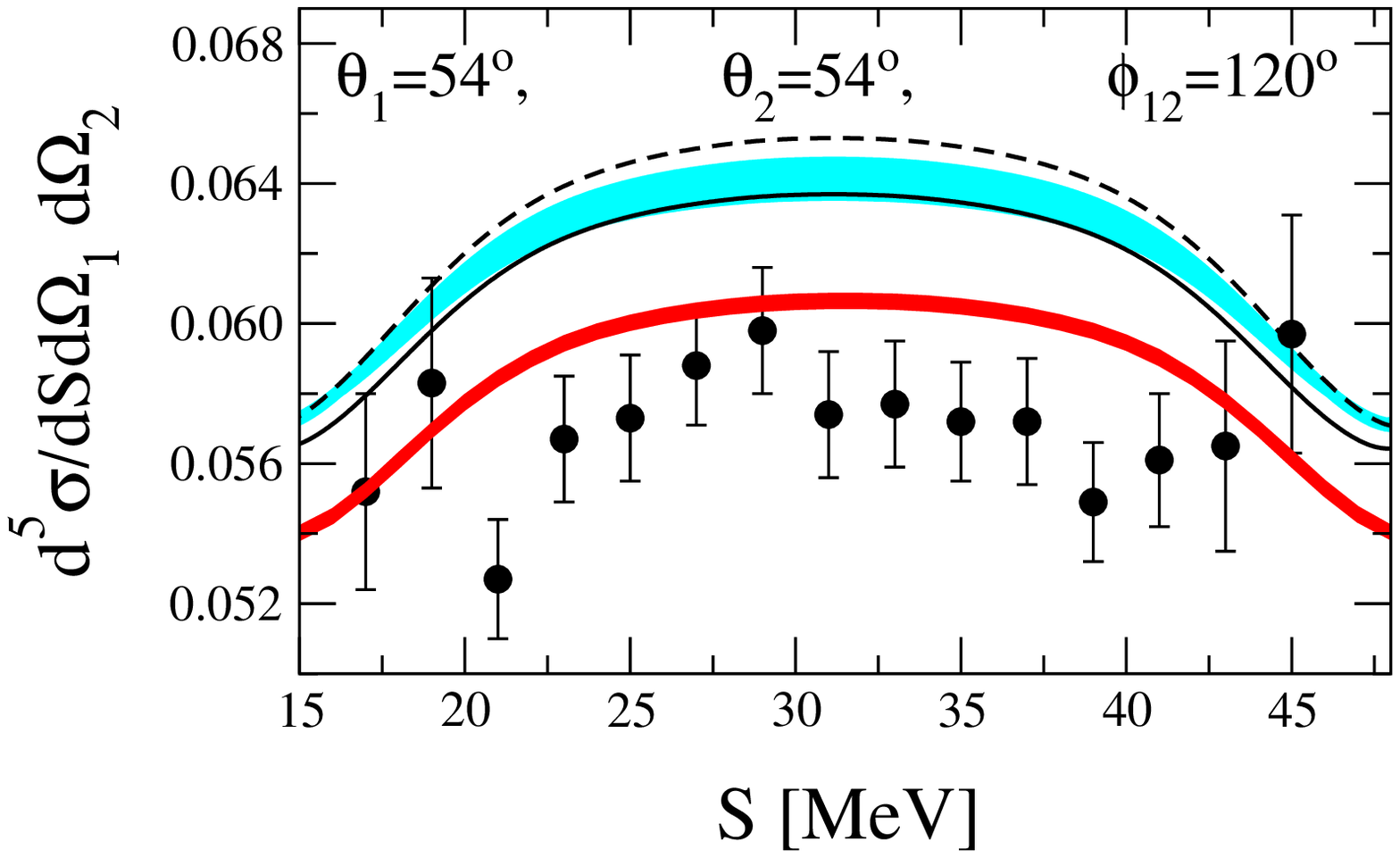,width=7.4cm}
\psfig{file=ay65.epsi,width=6.3cm}
}
\vskip -0.7 true cm
\centerline{
\parbox{1.00\textwidth}{
\caption[fig3N2]{
\label{fig3N2}  
Left: $pd$ breakup cross section data in [mb MeV$^{-1}$ sr$^{-2}$]
along the kinematical locus S (in MeV) at 65 MeV. The data are from \cite{zejma}.
NNLO predictions (dark shaded band) and NLO  
(light shaded band) compared to the conventional 
NN forces$+$TM 3NF predictions. The solid (dashed) line refers to the 
AV18$+$URBANA IX (CD Bonn$+$TM') results. Right: Analyzing power.
}}}
\end{figure}

\noindent 
In Fig.~\ref{fig3N3}, the tensor analyzing powers $T_{20}$ and $T_{22}$
for elastic $pd$ scattering at 70~MeV are shown \cite{seki,sekipriv}. While  $T_{20}$
is well described by the NNLO calculation, one observes some visible differences
in $T_{22}$. It is conceivable that this will be cured by the N$^3$LO 3NFs, which
are presently worked out. We believe that these corrections to the 3NF are free
of LECs, that means even at  N$^3$LO one has only two parameters to completely
pin down the chiral 3NF.
\noindent There is also a large amount of $pd$ break-up data, for a detailed
comparison between theory and experiment, see \cite{EGM3NF}. For illustration,
in Fig.~\ref{fig3N2} the break-up cross section in the symmetric space-star
configuration as well as the analyzing power are shown, in comparison to the
conventional approach based on two high-precision $NN$ potentials and two
available phenomenological 3NFs.
\noindent In \cite{EGM3NF}, we also solved the Yakubowsky equations to determine the
binding energy (BE) of the $\alpha$-particle. One finds (all numbers in MeV):
\beq
{\rm NNLO:}~~ {\rm BE}(^4{\rm He}) = -29.98 \ldots -29.51~, \quad 
{\rm Exp.:}~~{\rm BE}(^4{\rm He}) = -29.8\pm 0.1~,
\eeq
where the experimental value is the ``synthetic'' binding energy for pure $np$
forces to allow for a direct comparison with the then available chiral EFT forces
(for details see \cite{EGM3NF}). There  has also been some more 
recent work on the isospin
dependence of the three--nucleon force   and its effect on the binding in the
3-nucleon system \cite{EGMiso,FPvK}.

\section{RESONANCE SATURATION OF FOUR-NUCLEON COUPLINGS}

\begin{figure}[htb]
\begin{minipage}{6.2cm}
\begin{center}
\psfig{file=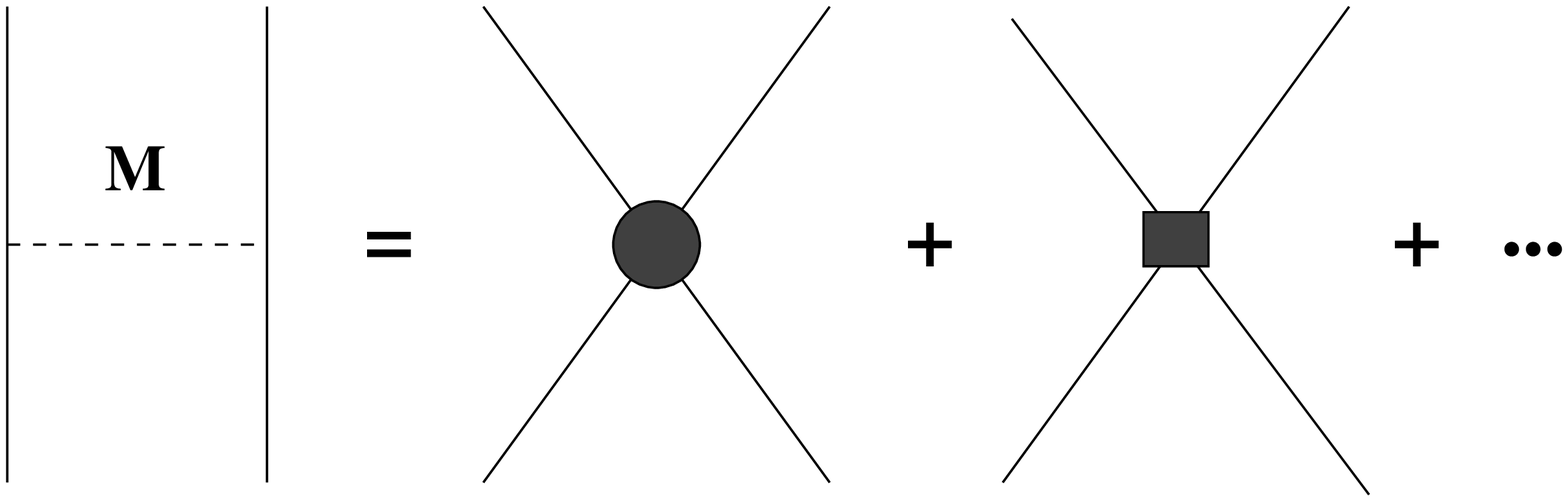,width=5.cm}\\[0.3em]
\end{center}
Figure~8: Top: Heavy meson exchange represented
as a sun of local interactions with increasing number
of derivatives. Right: Comparison of the NLO (green bands)
and NNLO (red bands) LECs with values obtained from 
accurate potentials \cite{EMGE}.
\end{minipage}
\hskip 1.7 true cm
\begin{minipage}{8.5cm}
\psfig{file=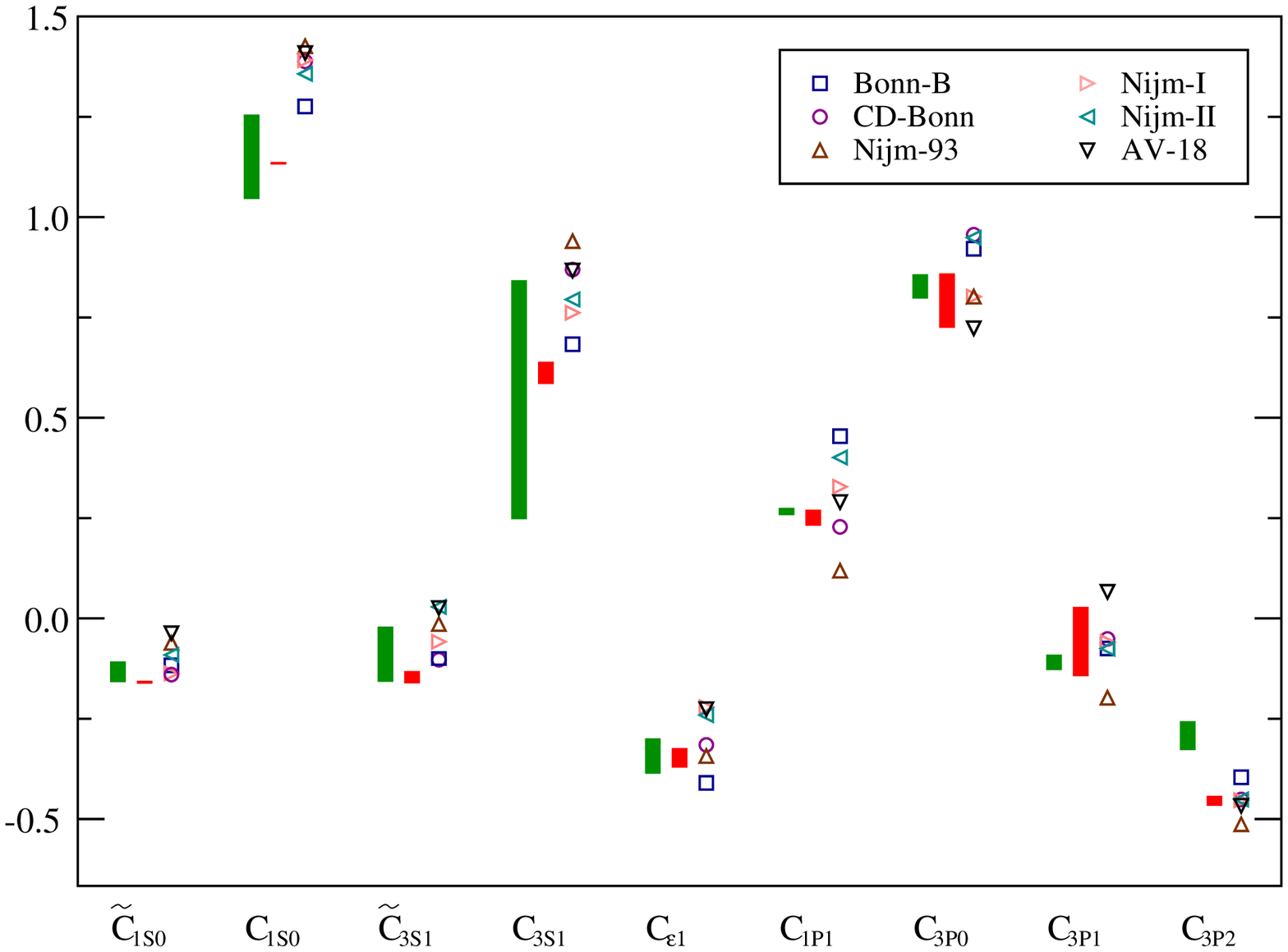,width=8.2cm}
\end{minipage}
\label{figreso}
\end{figure}

In the meson sector of CHPT  it has been demonstrated that the numerical 
values of the pertinent NLO LECs $L_i$ can  be understood to a 
high precision by integrating out heavy 
mesons of all types from the theory \cite{Eetal,DRV}. 
This was coined resonance saturation.
For the dimension two and and some dimension three LECs of the pion-nucleon
sector, these ideas were extended in~\cite{BKMlec} and later used in
studies of neutral pion electroproducion off protons.
It thus appears natural to confront the four-nucleon  LECs
determined from nuclear EFT with the highly
successful phenomenological/meson exchange models of the nuclear force
following the lines of Ref.~\cite{EMGE} and extending the
ideas of resonance saturation to this sector.
To be specific, consider some genuine one--boson--exchange (OBE) models.
In these models  the long range part of the interaction is given
by OPE (including a pion--nucleon form factor) whereas shorter
distance physics is expressed as a sum over heavier meson-exchange 
contributions:
\beq
V_{\rm NN} = V_\pi + \sum_{M=\sigma, \rho, \ldots} V_M~.
\eeq
Here some mesons can be linked to real resonances (like e.g. the
$\rho$--meson) or are parameterizations of certain physical effects, e.g. the
light scalar--isoscalar $\sigma$--meson is needed to supply the intermediate
range attraction (but it is {\em not} a resonance).
The corresponding meson--nucleon vertices are given
in terms of one (or two) coupling constant(s) and corresponding form factor(s),
characterized by some cut--off scale. These form factors are needed to
regularize the potential at small distances (large momenta) but they
should not be given a physical interpretation. 
As depicted in Fig.\ref{figreso} (left panel)
for nucleon momentum transfer below the masses of the exchanged mesons,
one can interpret such exchange diagrams as a sum of local operators
with increasing number of derivatives (momentum insertions).
This is explained in detail in Ref.~\cite{EMGE}.
In that work the short--range part
of different phenomenological potential models was power expanded 
and the resulting 
contact operators  were compared with the ones in the EFT approach. 
The latter have
to be corrected by adding the corresponding power expanded TPE contributions, 
which are not present in most phenomenological models.   
It was then demonstrated explicitly that the values 
of the LECs $C_i$ determined from various phenomenological OBE 
models are close to the values found in EFT at NLO and NNLO, see
Fig.~\ref{figreso}.  This was repeated in \cite{EGMspec} utilizing 
spectral function regularization with a similar outcome - the LECs 
appearing in the chiral EFT fits can be well represented by heavy meson
exchanges.
\setcounter{figure}{8}

\section{QUARK MASS DEPENDENCE OF THE NUCLEAR FORCES}
\begin{figure}[htb]
\centerline{
\psfig{file=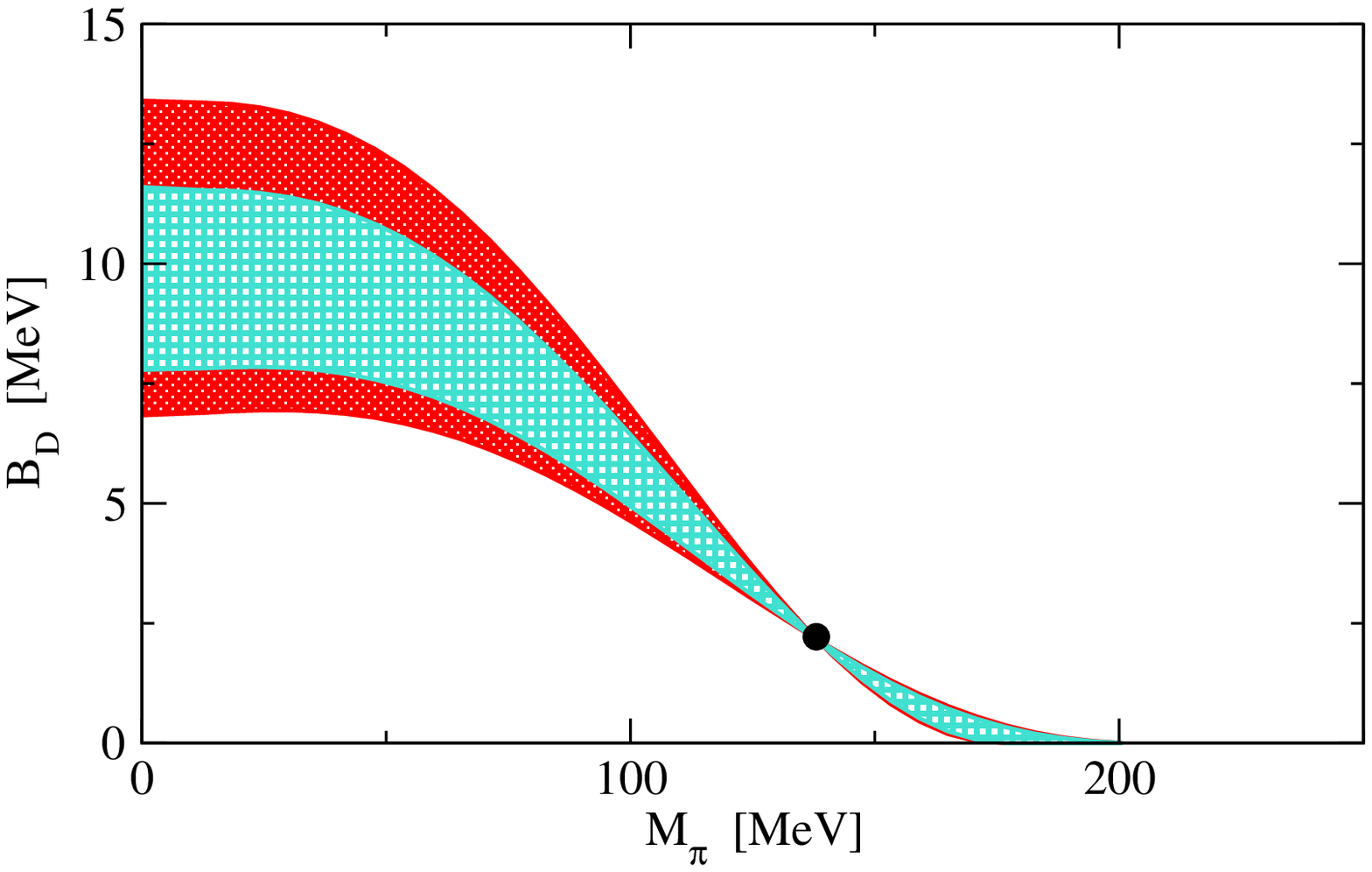,width=9.cm}
\psfig{file=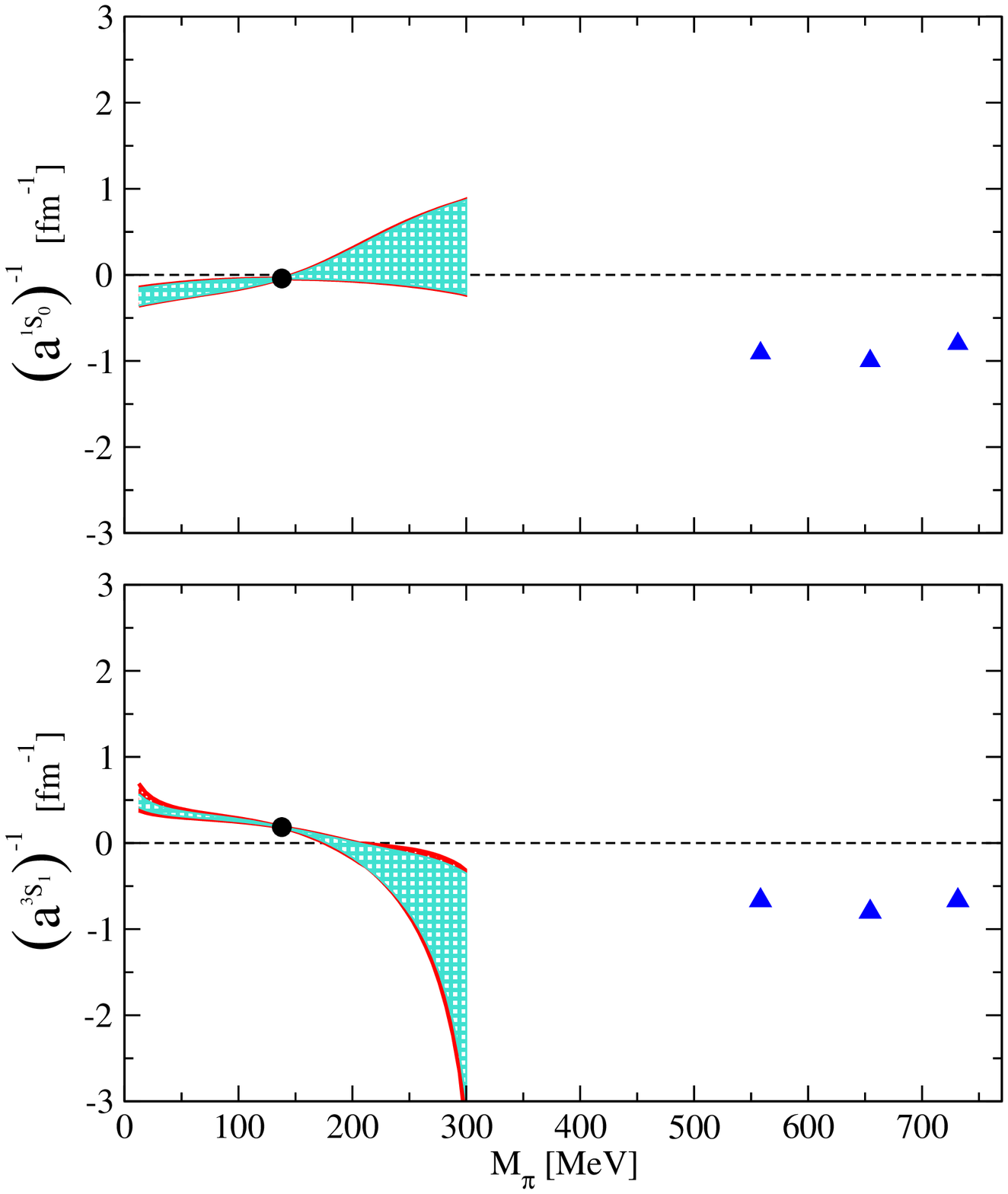,width=5.4cm}
}
\vskip -0.7 true cm
\centerline{
\parbox{1.00\textwidth}{
\caption[figl]{
\label{figcl}
Left panel: Deuteron BE versus the pion mass. The shaded areas show 
allowed values. The light shaded band corresponds to our main result with  
the uncertainty due to the unknown LECs $\bar D_{S,T}$.
The dark shaded band gives the additional uncertainty due to the 
uncertainty of $\bar d_{16}$.
The heavy dot shows the BE for the physical case $\tilde M_\pi = M_\pi$
Right panel: The inverse S--wave scattering lengths as functions of $\tilde M_\pi$.
The shaded areas represent the allowed values according to our analysis.
The heavy dots corresponds to the values in the real world. The triangles
refer to lattice QCD results from \cite{Fukug95}.
}}}
\end{figure}

Because of the smallness of the up and down quark masses, one does not expect
significant changes in systems of pions or pions and one nucleon when the
quark masses are set to zero (with the exception of well understood chiral
singularities like e.g. in the pion radius or the nucleon
polarizabilities). The situation is more complicated for systems of two (or
more) nucleons, as first discussed in EFT in \cite{BBSvK}. Here, I report on
similar work \cite{EMG} that is mostly concerned with the properties of the
deuteron and the S-wave scattering lengths as a function of the quark (pion)
mass. These questions are not only of academic interest, but also of practical
use for interpolating results from lattice gauge theory. E.g. the S-wave
scattering lengths have been calculated on the lattice using the quenched 
approximation \cite{Fukug95}. Another interesting application is related to
imposing bounds on the time-dependence of some fundamental coupling constants
from the NN sector, as discussed in \cite{Beane02}. To address this issue,
at NLO the following contributions have to be
accounted for (in addition to the LO OPE and contact terms without derivatives): 
i) contact terms with two derivatives or one $M_\pi^2$--insertion, 
ii) renormalization of the OPE,
iii) renormalization of the contact terms, and
iv) two--pion exchange (TPE).
This induces {\em explicit} and  {\em implicit} quark mass dependences. In the
first category are the pion propagator that becomes Coulomb-like in the
chiral limit or the $M_\pi^2$ corrections to the leading contact terms. These
are parameterized by the LECs $\bar D_{S,T}$ at NLO. These LECs can at present 
only be estimated using dimensional analysis and resonance saturation \cite{EMGE}. 
The implicit pion mass dependence enters at NLO through the pion--nucleon
coupling constant (note that the quark mass dependence of the nucleon
mass only enters at NNLO) expressed through the pion mass dependence of $g_A/F_\pi$
in terms of the quantity
\beq
\label{deltaCL}
\Delta = \left( \frac{g_A^2 }{16 \pi^2 F_\pi^2} - \frac{4 }{g_A}
\bar{d}_{16} + \frac{1}{16 \pi^2 F_\pi^2} \bar{l}_4 \right) 
(M_\pi^2 - \tilde M_\pi^2) - 
\frac{g_A^2 \tilde M_\pi^2}{4 \pi^2 F_\pi^2} \ln \frac{\tilde M_\pi}{M_\pi} \, .
\eeq
Here $\bar l_4$, $\bar d_{18}$ and $\bar d_{16}$ are LECs related to 
pion and pion--nucleon interactions, and the value of the pion mass is denoted
by $\tilde M_\pi$ in order to distinguish it from the physical one denoted by
$M_\pi$. In particular, $\bar d_{16}$ has been determined in various fits to
describe $\pi N \to \pi\pi N$ data, see \cite{FBM}.
The deuteron BE as a function of the pion mass is shown in Fig.\ref{figcl},
we find that the deuteron is stronger bound in the chiral limit (CL) than in the
real world,
\beq
B_{\rm D}^{\rm CL} =  9.6 \pm 1.9 {{+ 1.8} \atop  {-1.0}}~{\rm  MeV}~,
\eeq
where the the  
first indicated error refers to the uncertainty in the value of $\bar D_{^3S_1}$ and 
$\bar d_{16}$ being set to its average value 
while the second indicated error shows the additional uncertainty due 
to the uncertainty in the determination of $\bar d_{16}$
\noindent We find no other bound states, although the higher $S=1$ partial waves
rise linear with momentum due to the Coulomb-like pion propagator.
Last but not least,
we found smaller (in magnitude) and more natural values for the two 
S--wave scattering lengths in the chiral limit,
\beq
a_{\rm CL} (^1S_0) = -4.1 \pm 1.6 {{+ 0.0} \atop 
{-0.4}} \,{\rm fm}~, \quad {\rm   and} \quad
a_{\rm CL} (^3S_1) = 1.5 \pm 0.4 {{+ 0.2} \atop \\ 
{ -0.3}}\, {\rm fm}\, .
\eeq
As stressed in \cite{EMG}, one needs lattice data for pion masses below
300 MeV to perform a stable interpolation to the physical value of $M_\pi$,
cf the right panel in Fig.~\ref{figcl}.
We conclude that nuclear physics in the chiral limit is much more natural than 
in the real world. Another interesting application of the quark mass of the
nuclear forces is the recently conjectured infrared renormalization group
limit cycle in the three--nucleon system \cite{BH}.

\section{BRIEF SUMMARY AND OUTLOOK}

From the previous sections it should have become obvious that few-nucleon systems can be 
studied in chiral effective field theory in a systematic and model-independent
way. The two-nucleon system has been analyzed at N$^3$LO and accurate results
for the deuteron and scattering observables have been obtained. Furthermore,
3N, 4N and even 6N systems \cite{nocore} have been studied at N$^2$LO. For the
first time, the chiral three--nucleon force has been consistently included and the
results to this order look very promising. Many other applications have also been
performed (some times using the so-called hybrid approach which utilizes the kernel 
from EFT and wave functions from semi-phenomenological potentials). I mention just 
a few here: pion-deuteron scattering \cite{pid}, neutral pion photo- and electroproduction
on the deuteron \cite{photo}\cite{electro}, Compton scattering off deuterium 
\cite{compton}, nuclear
parity violation \cite{pv} or solar fusion and the $hep$ process \cite{hep}.
Other interesting developments are related to the lattice, just to name a few examples
of recent work:  nuclear lattice simulations \cite{dean}, the two-nucleon potential
in partially quenched lattice QCD \cite{sima} or the discussion of the size of the 
lattice required to simulate the two-nucleon system \cite{BBPS}.

\bigskip
\noindent
From my point of view, the following problems should be worked out next:
\begin{itemize}
\item To study few-nucleon physics at N$^3$LO, we have to work out the
3NF and 4NF at this order. It is gratifying to notice that no new six-nucleon
contact interactions appear at this order and thus one has large predictive
power. Furthermore, the  4NF first appears at this order and is  parameter-free.
It is expected that this force will be much smaller than the 3NF. Still, it
will be interesting to obtain the quantitative size of this force.
\item Similarly, the currents corresponding to electroweak probes
have to be constructed to the same order. The strength of the chiral EFT
approach is the consistency with the forces and the automatic incorporation
of gauge symmetry. Lots of work on this problem has already been done by the
Korean group  \cite{parketal} (see also \cite{walzl,drp}). However, most of these
results need to be rederived using the method of unitary transformation to make them 
consistent with the forces. Some pertinent diagrams are shown in Fig.~10.
Furthermore, three--nucleon currents have never been
considered.
\end{itemize}

\setcounter{figure}{10}

\begin{figure}[tb]
\begin{minipage}{5.2cm}
Figure~10: $2N$ and $3N$ currents. First row:
Leading pion exchange graphs. Second row: Corrections
of various ranges. Third row: $3N$ currents. Solid, dashed,
and wiggly lines represent nucleons, pions and photons, in order.
Only some representative diagrams are depicted.
\end{minipage}
\hskip 1.7 true cm
\begin{minipage}{8.5cm}
\psfig{file=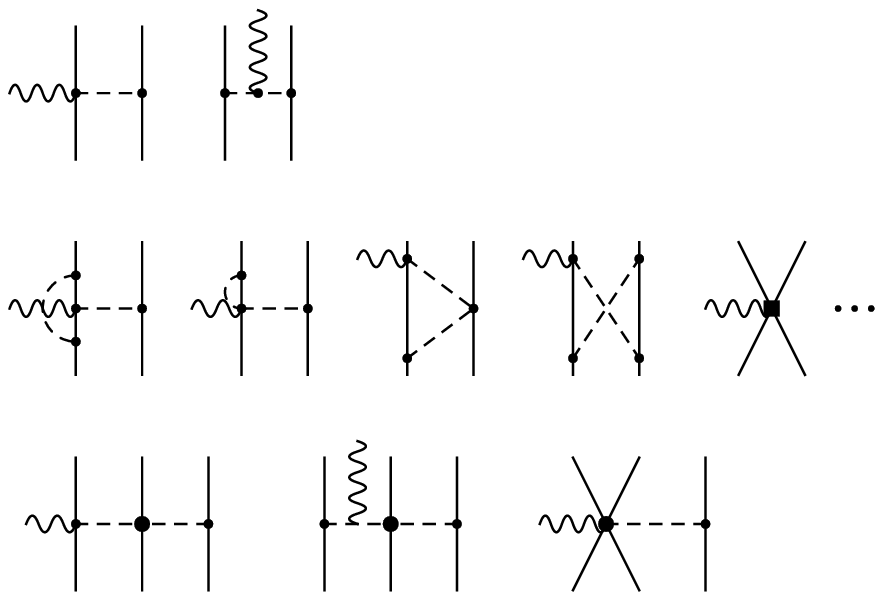,width=7.7cm}
\end{minipage}
\label{figcurr}
\end{figure}

\noindent The methods described here  pave the way to a precision nuclear physics
consistent with the symmetries of QCD and thus offer a sound theoretical
foundation for this decade old problem.

\section*{ACKNOWLEDGMENTS}

It is a pleasure to thank my collaborators on these topics, and particularly 
Evgeny Epelbaum, Walter Gl\"ockle, Hiroyuki Kamada, Andreas Nogga and  
Henryk Wita{\l}a. I am grateful to Evgeny Epelbaum for a careful reading of this
manuscript. I also would like to thank the organizers, in particular Bj\"orn
Jonson, for the invitation and the excellent organization.

\end{document}